\documentclass[%
 aip,
 apl,
 amsmath,amssymb,
 reprint,%
]{revtex4-1}

\usepackage{graphicx}% Include figure files
\usepackage{dcolumn}% Align table columns on decimal point
\usepackage{bm}% bold math
\usepackage[utf8]{inputenc}
\usepackage[T1]{fontenc}
\usepackage{mathptmx}
\usepackage{etoolbox}
\usepackage{color}
\usepackage{xcolor}
\usepackage{graphicx}

\newcommand{\params}[0]{\mathbf{\theta}}
\newcommand{\model}[0]{\mathcal{M}}

\makeatletter
\def\@email#1#2{%
 \endgroup
 \patchcmd{\titleblock@produce}
  {\frontmatter@RRAPformat}
  {\frontmatter@RRAPformat{\produce@RRAP{*#1\href{mailto:#2}{#2}}}\frontmatter@RRAPformat}
  {}{}
}%
\makeatother
\begin{document}

\preprint{AIP/123-QED}

\title[Bayesian modelling of scattered light]{Bayesian modelling of scattered light in the LIGO interferometers}
% Force line breaks with \\
\author{R. P. Udall}
\author{D. Davis}%
 \email{rudall@caltech.edu.}
\affiliation{$^{1)}$
LIGO, California Institute of Technology, Pasadena, CA 91125, USA}%

\date{\today}% It is always \today, today,
             %  but any date may be explicitly specified

\begin{abstract}
Excess noise from scattered light poses a persistent challenge in the analysis of data from gravitational wave detectors such as LIGO. We integrate a physically motivated model for the behavior of these ``glitches'' into a standard Bayesian analysis pipeline used in gravitational wave science. This allows for the inference of the free parameters in this model, and subtraction of these models to produce glitch-free versions of the data. We show that this inference is an effective discriminator of the presence of the features of these glitches, even when those features may not be discernible in standard visualizations of the data. 
\end{abstract}

\maketitle

%\textbf{Requirements:\\}
%\textbf{At the time of submission, your paper must be 3000 words or less, including figure and table captions. This word count excludes the title, author list, abstract, acknowledgements, data availability statement, references, and non-text items including figures, tables, and equations. Normally, papers longer than 3000 words will be rejected, and the authors will be required to shorten the paper and submit it again as a new manuscript. Please also note that section headings are not permitted for letters.}

%\textbf{Authors are encouraged to include up to five figures in the paper to illustrate their results. Images may be grouped together as separate panels in one figure if they are logically connected and clarity is not compromised.}

%\textbf{The sections listed in this draft right now will need to be deleted and our just for our own benefit.}

%\subsection{Introduction}

The Laser Interferometer Gravitational-wave Observatory (LIGO) includes a pair of ground-based interferometers. LIGO Hanford and LIGO Livingston, designed to be sensitive to gravitational waves from astrophysical sources~\cite{aligo}.
The LIGO detectors are dual-recycled Michelson interferometers with Fabry-Perot arm cavities 4 kilometers in length.
Recent observations alongside the Virgo~\cite{avirgo} and KAGRA~\cite{kagra} detectors have identified 90 gravitational-wave signals from the mergers of compact objects~\cite{GWTC-2,GWTC-3,GWTC-2.1}.

The sensitivity of LIGO is limited both by persistent noise sources, such as quantum noise and thermal noise~\cite{aLIGO:2020wna,avirgo}, as well as transient noise sources, such as earthquakes and thunder~\cite{GW150914_detchar,LIGO:2021ppb,Virgo:2022fxr,AdvLIGO:2021oxw}. 
Transient noise is manifested in detector data as short-duration bursts of excess power, commonly referred to as ``glitches.''
Glitches can create challenges for gravitational-wave analyses by either mimicking gravitational-wave signals (preventing identification of signals~\cite{TheLIGOScientific:2017lwt,Nitz:2017lco,Cabero:2019orq,Davis:2020nyf,LIGO:2021ppb,Ashton:2021tvz}) or by overlapping detected events (preventing further analysis~\cite{Pankow:2018qpo,Powell:2018csz,Macas:2022afm, Kwok:2021zny, Davis:2022ird}).

One of the most common sources of transient noise in recent observing runs is from ``scattered light''~\cite{Soni:2020rbu, Soni:2021cjy}.
Imperfections in the mirror surfaces can lead to light scattered off of the main beam path in the interferometer. 
If the light reflects off of a moving object and then rejoins the main beam path, this adds an additional phase shift to the observed light.
Although scattered light may occur randomly, the impact of the source of reflection is well understood. If the scattering occurs from one of the test masses, the most likely source of reflection is the suspension system supporting the test mass or nearby optical components. 

\begin{figure*}[t]
\includegraphics[width=0.6\textwidth]{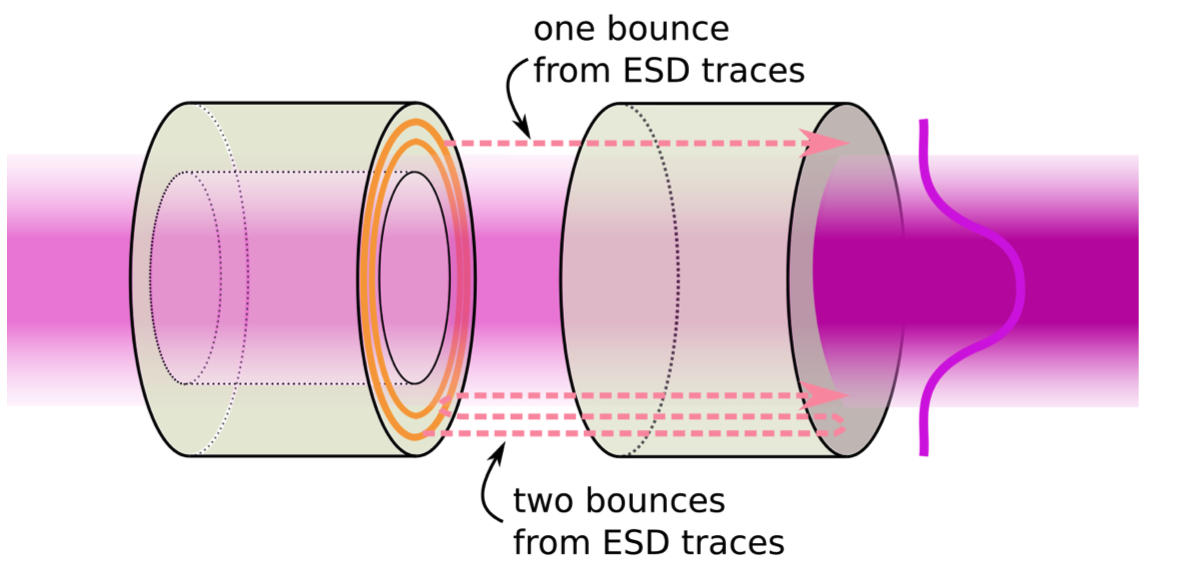}
\caption{\label{fig:suspensions} A diagram of the test mass and reaction chain mass showing the path that of scattered light. Some light is transmitted through the test mass and reflects off components of the electrostatic drive (ESD). As the direction of reflection is parallel with the main beam path, it is possible for the light to be reflected multiple times before rejoining the main beam path. Figure is reproduced from \emph{Soni et al.}~\cite{Soni:2020rbu}}
\end{figure*}

The most common source of reflection for scattered light in LIGO during the third recent observing run (O3) was from light reflected off of the reaction reaction mass that is located directly behind the test masses at the end of each arm of the interferometer~\cite{Soni:2020rbu}. 
Figure~\ref{fig:suspensions} shows the optical path of the scattered light from this source of reflection.
The reaction chain mass contains an electrostatic drive (ESD) that is used to control the motion of the main test mass. 
The highly reflective gold coatings used as part of the ESD allowed scattered light to reflect a large number of times, creating scattered light glitches with many arches. 
The rate of scattered light light glitches from this source and others was so high that $\approx20\%$ of gravitational-wave signals in O3 overlapped in time with such glitches~\cite{GWTC-2,GWTC-3}. 

In order to prevent glitches present in the data from biasing estimates of the source properties of gravitational-wave signals, any glitches that are nearby or directly overlap gravitational-wave signals are subtracted from the data~\cite{Pankow:2018qpo, Davis:2022ird}. 
Using information from the motion of the suspension system, one can estimate the frequencies of scattered light glitches~\cite{Accadia:2010zzb,VIRGO:2011tzh,Valdes:2017xce,Bianchi:2021unp, Longo:2021avq, gwdetchar}. However, since these monitors only approximate the motion of the specific surface that is reflecting the light, it is difficult to precisely model individual scattered light glitches using these techniques.
Other methods exist to subtract scattered light glitches, but these tools either use weakly modelled techniques~\cite{Cornish:2014kda,Pankow:2018qpo,Cornish:2020dwh, Hourihane:2022doe, Ashton:2022ztk} or require that there is a relevant sensor which accurately witnesses the source of the glitch~\cite{Davis:2019, Mogushi:2021deu,Davis:2022ird}.
As scattered light glitches have a relatively long duration, rapidly changing frequency, and only witnesses that approximate the true source of the glitching, modelling and subtracting these glitches with these additional techniques is challenging. 
The high rate of these glitches, combined with a well-understood model for the relevant instrumental mechanism that creates the glitches, makes scattered light glitches ideal candidates to instead model using an analytic methods.

In this letter, we demonstrate how we can utilize the simple harmonic motion of the Advanced LIGO suspension system to model scattered light glitches. 
We then use Bayesian inference to estimate the parameters of individual scattered light glitches. 
This tool can be used to precisely model scattered light glitches for both detector characterization and to improve the Bayesian inference of the source properties of detected gravitational-wave events.

%\begin{figure}[t]
%\includegraphics[width=\columnwidth]{Scattering_omega.pdf}
%\caption{\label{fig:soectrogram} Here is a scattered light glitch.}
%\end{figure}

%\subsection{Scattering model (Derek) }

Generically, the excess strain noise, $h(t)$, produced by scattering of light with wavelength $\lambda$ can be modeled based on the motion of the relevant surface, $x(t)$~\cite{Accadia:2010zzb}:
\begin{equation}
h(t) = A \sin\left[ \frac{4\pi}{\lambda} x(t) + \phi \right].
\end{equation}
Therefore, the instantaneous frequency of this strain noise is 
\begin{equation}\label{eq:scat_freq}
f(t) = \left| \frac{1}{2\pi} \frac{d}{dt} \left(\frac{4\pi}{\lambda} x(t) + \phi \right)  \right| 
= \left| 2\frac{v(t)}{\lambda} \right|
\end{equation}
This equation can be used to predict the frequency of the related scattered light glitches. 
If the scattered light is reflected $N$ times, the scattered light glitch frequency is $N$ times the frequency predicted by Equation~\ref{eq:scat_freq}.
An example of LIGO data containing a scattered light glitch and the predicted glitch frequencies can be seen in Figure~\ref{eq:scat_freq}.
This relationship has been used extensively to help identify the source of scattered light that is observed in gravitational-wave interferometers~\cite{Accadia:2010zzb,VIRGO:2011tzh,Valdes:2017xce,Bianchi:2021unp, Longo:2021avq, gwdetchar}; 
if the motion of a component in the detector predicts the same frequency evolution as is observed in the strain data, it is likely that the source is nearby this component.

If we assume that the surface that reflects the scattered light is a simple harmonic oscillator with frequency $f_{\text{mod}}$, then the motion of the optic is 
\begin{equation}
x(t) = C \sin\left( 2\pi f_{\text{mod}} t \right).
\end{equation}
and then setting $f_{\text{harm}} = \text{max} [f(t)] = \frac{4\pi C}{\lambda} f_{\text{mod}}$, we arrive at our scattering equation:~\cite{Tolley}
\begin{equation}
h(t) = A \sin\left[ \frac{f_{\text{harm}}}{f_{\text{mod}}}\sin\left( 2\pi f_{\text{mod}} t \right) + \phi \right].
\end{equation}
Physically, $f_{\text{harm}}$ corresponds to the maximum frequency of the scattered light glitch, $1/f_{\text{mod}}$ is twice the duration of each glitch, and $A$ is the amplitude of the strain induced by the glitch.
Typically, the relevant frequencies of seismic motion responsible for scattered light glitches are in the microseismic ($\approx$1/6\,Hz) band, resulting in a scattered light glitch every 3 seconds and $f_{\text{mod}}\approx1/6$\,Hz.

\begin{figure*}[t]
\includegraphics[width=0.75\textwidth]{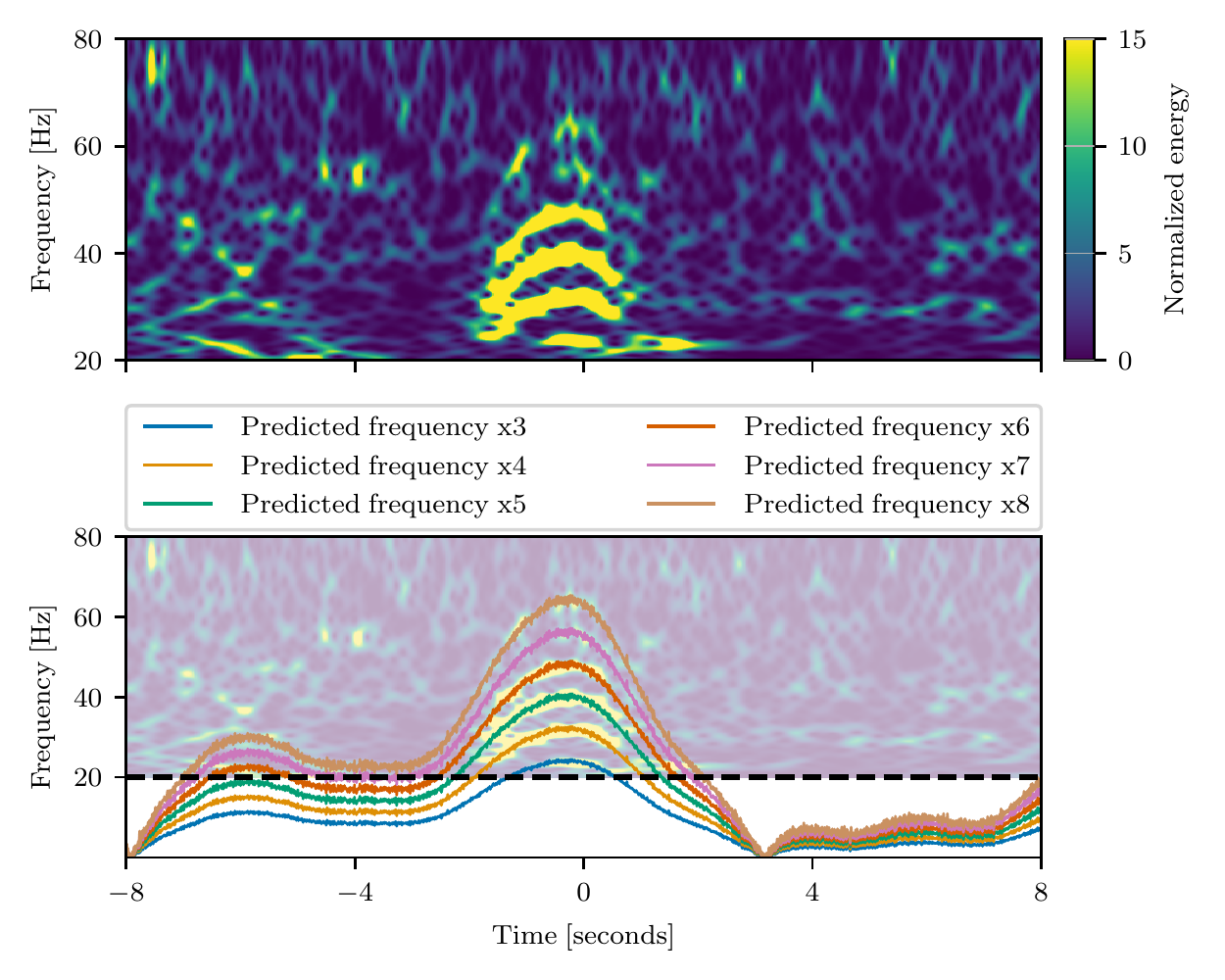}
\caption{\label{fig:scat_track} A comparison of data containing a scattered light glitch and the suspension motion data that is correlated with the glitch morphology. The top shows a spectrogram of data from LIGO Livingston during the second observing run showing an isolated scattered light glitch with many harmonics. The bottom shows the same spectrogram with lines overlaid showing the frequency of the scattering arches as predicted by the motion of the second stage of the suspension system holding one of the test masses. The dashed black line marks the minimum frequency of the spectrogram data. The suspension motion data has been multiplied by 1.25 to better match the observed arches. With this correction, there is clear agreement with the predicted and true frequencies of the scattering arch. }
\end{figure*}

%%%% TWO FREQUENCY MODEL

%If we instead consider an oscillator with two frequencies, $f_{\text{mod},1}$ and $f_{\text{mod},2}$, then this equation is  
%\begin{equation}
%\begin{split}
%h(t) = A \sin\biggl[ &\frac{f_{\text{harm},1}}{f_{\text{mod},1}}\sin\left( 2\pi f_{\text{mod},1} t \right) + \\ &\frac{f_{\text{harm},2}}{f_{\text{mod},2}}\sin\left( 2\pi f_{\text{mod},2} t \right) + \phi \biggr].
%\end{split}
%\end{equation}
%In this case, the maximum frequency is $f_{\text{harm},1} + f_{\text{harm},2}$.
%The value of $A$ is dependent on the physical properties of the cavity of interest. 

%Typically, the two relevant frequencies are the and microseismic ($\approx$1/6\,Hz) and anthropogenic ($\approx$2\,Hz) bands. As $f_{\text{mod, an}} > f_{\text{mod, ms}}$, scattering due to both these frequencies can either be seen as arches with a rate of 4\,Hz (when $f_{\text{harm, an}} \gg f_{\text{harm, ms}}$) or 
%2\,Hz (when $f_{\text{hard, an}} \approx f_{\text{harm, ms}}$).
%Furthermore, the maximum frequency of 4\,Hz scattering is relatively stable, while the frequency of 2\,Hz scattering will change on the timescale of $f_{\text{mod, ms}}$.

%\subsection{Bayesian analysis}
To proceed, we now include the previosuly mentioned complication to this picture: scattering events consist of many harmonics, appearing on top of each other with increasing harmonic frequency and decreasing amplitude, and so accurately characterizing the glitch and removing it from the data stream requires inclusion of all of these harmonics.
The frequencies of each harmonic are integer multiples of each other, such that the frequency separation between any two adjacent harmonics is nearly constant; we denote this separation as $f_{diff}$.
Similarly, the modulation frequency and central time are almost exactly constant.
Accordingly, we model a collection of $N$ of arches using these relations, with perturbation terms to account for small variations between harmonics:
\begin{equation}
    h(t) = \sum_{k=0}^N A_k \sin\biggr{[}\frac{f_{harm, k}}{f_{mod, k}}\sin(2\pi f_{mod, k} (t-t_{c,k}) + \phi_k)\biggr{]}
\end{equation}
Where for $k>0$
\begin{equation}
    f_{harm,k} = f_{harm, 0} + k f_{diff} + \delta f_{harm, k}
\end{equation}
\begin{equation}
    f_{mod, k} = f_{mod, 0} + \delta f_{mod, k}
\end{equation}
\begin{equation}
    t_{c,k} = t_{c, 0} + \delta t_{c, k}
\end{equation}
This model greatly increases the sampling efficiency by explicitly including these relationships between harmonics, thus substantially reducing the configuration space of glitch morphology which must be searched.
Moreover, the relations in modulation frequencies and central times are consistently almost exact, such that we can set $\delta f_{mod, k} = \delta t_{c, k} = 0$ without affecting the quality of the subtraction, further reducing the complexity of the sampling problem. 
One limitation of this model is that the total number of harmonics must be fixed before beginning analysis of the data.
When we choose to fix the total number of harmonics to $N$, we describe this as the ``N~arch model.''
In a typical $N$ arch inference, we wish to determine the values of $3N +3$ parameters: the first arch is modeled with base parameters $A_0$, $f_{mod, 0}$, $f_{harm, 0}$, $t_{c,0}$, and $\phi_0$, while each subsequent arch is modeled with the shared difference parameter $f_{diff}$, as well as fully independent amplitudes $A_k$ and phases $\phi_k$, and harmonic frequency perturbation $\delta f_{harm, k}$.

This formula describes an infinite series of arches, and so we apply a Tukey window~\cite{Talbot:2021igi} of width $1 / 2f_{mod}$, and windowing parameter $\alpha=0.2$ centered about the arch sequence of interest.

Because this model has a large number of free parameters, we wish to infer posteriors on their true values, for which we turn to Bayesian inference~\cite{LIGOScientific:2019hgc}.
Bayesian inference is built upon Bayes theorem, which for data $d$, parameters $\params$, and model $\model$ is 
\begin{equation}
    p(\params | d, \model) = \frac{p(d|\params,\model)p(\params | \model)}{p(d | \model)}
\end{equation}
The model $\model$ consists of both the mapping from the parameters $\params$ to the expected data, and the a priori beliefs about the parameters $p(\params | \model)$.
More typically, we talk about the likelihood $\mathcal{L}(d | \params) \propto p(d | \params, \model)$, and the priors $\pi(\params)$.
Following convention, we will always notate posteriors as $p(\params)$ and priors as $\pi(\params)$, for brevity.
In this formulation we then have
\begin{equation}
    p(\params | d) = \frac{\mathcal{L}(d | \params) \pi(\params)}{Z}
\end{equation}
Where the evidence $Z$ is the normalizing factor, such that 
\begin{equation}
    Z = \int \mathcal{L}(d | \params) \pi(\params) \text{d} \params
\end{equation}
The likelihood for transient behaviors in the data stream, both glitches and true gravitational-wave signals, is well known under the assumption of Gaussian stationary noise~\cite{Ashton:2018jfp}, and is given by 
\begin{equation}
    \ln \mathcal{L}(\params) = -\frac{1}{2} \sum_k \biggr{[} \frac{|h_k(\params) - d_k|^2}{\sigma_k^2} + \ln(2\pi \sigma_k^2)\biggr{]} 
\end{equation}
Where $k$ denotes the frequency bin, $h_k(\params)$ is the value of the template for $\params$ at that frequency bin, $d_k$ is the value of the data in that frequency bin, and $\sigma_k$ is the value of the amplitude spectral density for the underlying noise in that frequency bin. 
Evaluation of the posterior must be done by numerical methods, typically by sampling with either Markov Chain Monte Carlo (MCMC)~\cite{MetropolisEtAl1953} or nested sampling methods~\cite{Skilling:2006}. 

\begin{figure*}[t]
\includegraphics[width=0.75\textwidth]{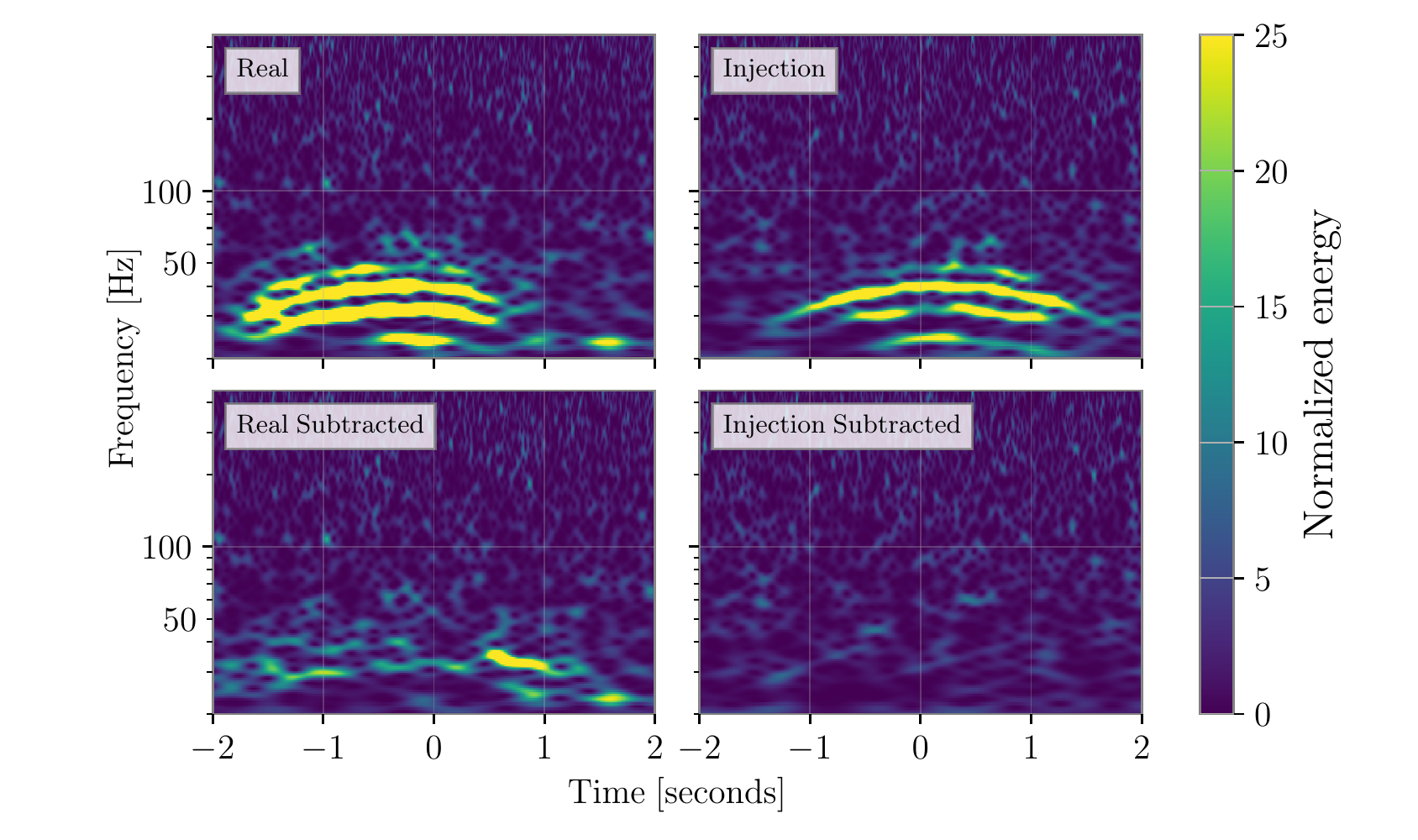}
\caption{\label{fig:subtract}
Here spectrograms are present for an example in data from LIGO Livingston during the second observing run (O2). 
The top left shows a real scattering event, corresponding to that plotted in Figure~\ref{fig:scat_track}, which have six arches visible in the data.
The bottom left shows the the subtraction of the reconstruction generated from maximum likelihood parameters for the inference of the scattering event, corresponding to the results plotted in the bottom left panel of Figure~\ref{fig:amplitude_histograms}.
Due to asymmetries in the scattering arch, as highlighted in Figure~\ref{fig:scat_track}, this subtraction is not perfect. 
Next, on the top right, we inject these maximum likelihood parameters into Gaussian noise, generated according to the amplitude spectral density inferred for a stretch of data from the second observing run (O2).
We then subtract the maximum likelihood parameters of the inference on this example, corresponding to the bottom right panel of Figure~\ref{fig:amplitude_histograms}.
Notably since the injection is perfectly symmetric, this subtraction is also highly effective.
}
\end{figure*}

We perform sampling with the interface provided by Bilby~\cite{Ashton:2018jfp}, a library for Bayesian inference which is specially adapted for use with Gravitational Wave data.
Bilby conveniently allows extension to custom models, and so we use this model in place of a standard gravitational-wave waveform template. 
Our sampling is done using the Dynesty sampling method~\cite{Speagle:2019ivv}, which is an implementation of the standard nested sampling method~\cite{Skilling:2006}. 
Briefly, nested sampling methods estimate the evidence by drawing points from the posterior, under a condition of monotonically increasing the minimum likelihood of the "live" set of points. 
The principal challenge of nested sampling is to draw these points efficiently, since for models with more than a few dimensions the volume of the typical set is many orders of magnitude smaller than the volume of the prior space, and so naive sampling would be woefully inefficient.
To do this more efficiently, we use differential evolution methods 
%experimentally included in Bilby\rhiannon{TODO figure out attribution for Colm}\derek{Do we know how to clean this up?},
which draw new proposed points using a set of multiple proposals, each using information about the ensemble of live points at that time. 
Furthermore, it is desirable to choose priors which are sufficiently constrained as to not waste computational time, while also being sufficiently uninformed as to not bias the inference. 
Since the frequencies and central times may be roughly estimated by eye from spectrograms, we set uniform priors centered upon these estimates for $f_{harm,0}$, $f_{mod,0}$, $f_diff$, and $t_{c,0}$.
Perturbations are centered on 0 with moderate width (for $\delta f_{harm, k}$, this width must be constrained to be less than $f_diff$, lest accidental degeneracies be introduced.  
Phases are estimated with a uniform prior over angles $[0, 2\pi)$, and amplitudes are estimated with log-uniform priors over 3 to 4 orders of magnitude, since their magnitudes are difficult to estimate from spectrograms. 

The wall time of these analyses vary depending on the configuration, the amplitude of the glitch, and the complexity of the parameter space being explored, but are consistently less than one day under reasonable conditions.
When run using a pool of 8 CPUs, the shortest run took approximately 20 minutes, while the longest took approximately 13 hours, which is typical for Bayesian inference problems with high dimensionality.

%\subsection{Injection tests and single arches}

To begin model validation, we conducted 4 injection tests in Gaussian noise, drawn from the amplitude spectral density of the data surrounding the isolated scattering event analyzed below.
These consisted of: 
\begin{enumerate}
    \item Injecting a scattered light glitch with 3 arches and recovering them with a 3 arch model,
    \item Injecting a scattered light glitch with 3 arches and recovering them with an 8 arch model,
    \item Injecting a scattered light glitch with 8 arches and recovering them with a 3 arch model,
    \item Injecting a scattered light glitch with 8 arches and recovering them with an 8 arch model.
\end{enumerate}

\begin{figure*}[t]
\includegraphics[width=0.9\textwidth]{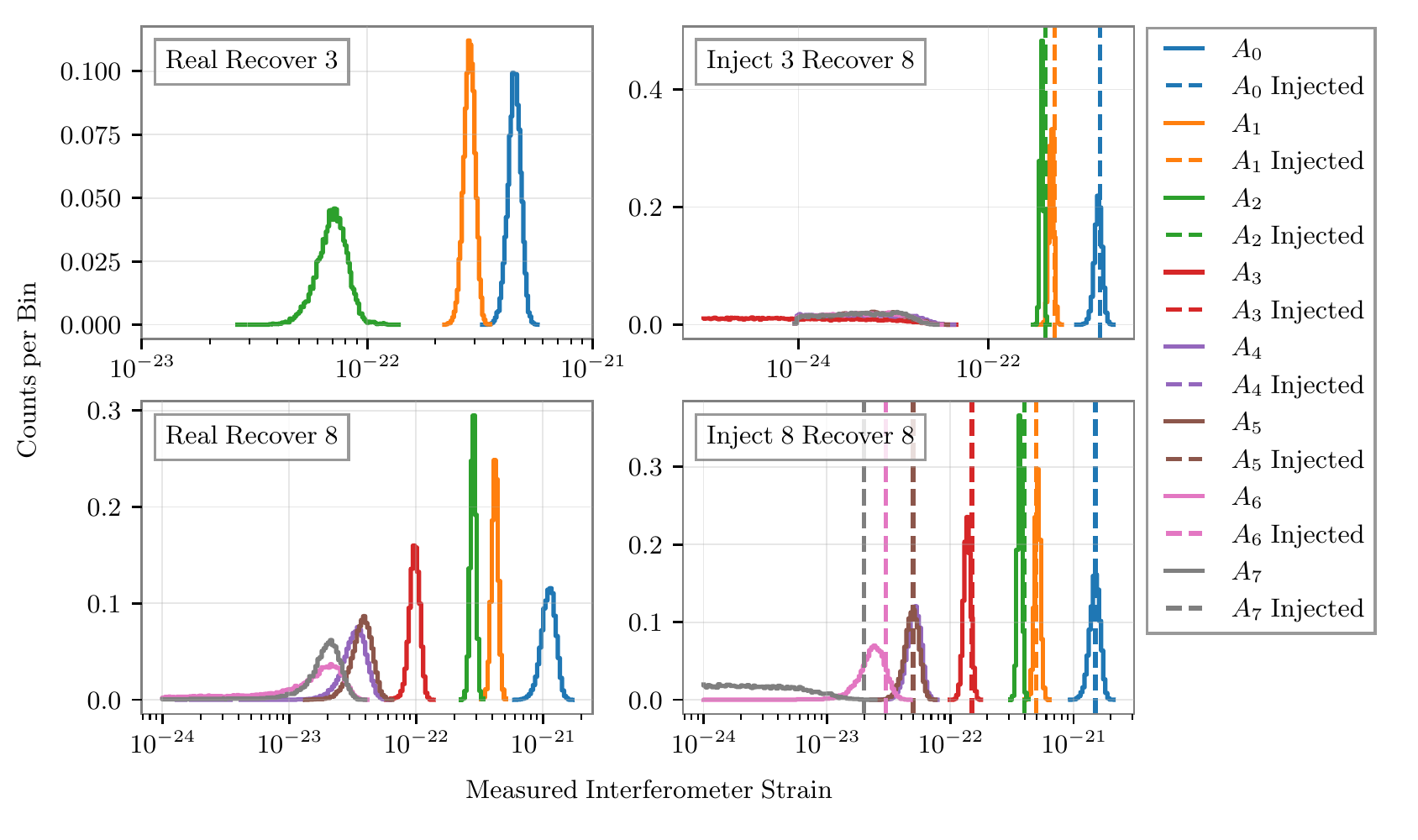}
\caption{\label{fig:amplitude_histograms}
Amplitude posteriors for four cases with O2 data and O2-like data are plotted.
The top left shows analysis of the O2 scattering example discussed, with inference done using a three arch model.
The bottom left shows the sample example, with inference done by an eight arch model.
Posteriors for arches recovered in both cases are shown to be consistent, while the eight arch model also recovers amplitudes constrained away from zero for all eight modelled arches, despite only six arches being visible in the corresponding spectrogram (Figure~\ref{fig:subtract}).
Subtraction of the maximum likelihood parameters for this panel may be seen in the bottom left panel of that figure.
On the right, we perform consistency tests by injecting maximum likelihood parameters for the signal analyzed on the left into Gaussian noise drawn from the power spectral density of the example data segment.
To test for the possibility of our model returning false positives, we first inject only three arches, while attempting to recover all eight, in the top right panel.
Notably, the five not injected have amplitude posteriors consistent with zero, indicating that they are recovering true negatives.
Finally, in the bottom left panel we inject eight arches and recover them with an eight arch model.
This actually returns a false negative for the eighth arch, which may be due to the specific noise realization, but does imply that the support for the eight arch in the real case must be quite strong. 
This injection corresponds to the top right panel of Figure~\ref{fig:subtract}, while the subtraction of these maximum likelihood parameters may be seen in the bottom right panel of that figure. 
}
\end{figure*}

The left column of Figure~\ref{fig:subtract} shows the injection of case 4, and its residual after subtraction of the model with the maximum likelihood parameters.
These tests should allow verification of a number of important points. Firstly, we should be able to accurately recover injected values, even when analyzing only a subset of the arches (case three).
Secondly, we should be able to distinguish the absence or presence of arches correctly, in cases two and four, respectively. 
Finally, we should be able to subtract the recovered parameters and arrive at nearly Gaussian noise. 
Figure~\ref{fig:amplitude_histograms} allows for a visualization of the accuracy of our recoveries and the distinguishability of the null case.
True (injected) values are accurately recovered for all injected arches. 
Furthermore, when only three arches are injected, recovery of the other five finds amplitudes that push toward zero, whereas when eight arches are injected, each arch except the last is recovered at distinctly nonzero values. 
Thus, when applied to true data, we have a meaningful test for model validity, when varying the number of arches being modeled.
For the eight arch case, this conclusion is further supported by the spectrograms in Figure~\ref{fig:subtract}, where the result of the subtraction is data which is visually indistinguishable from Gaussian noise. 

Next, we performed analysis of an isolated scattering event in O2 data, under both the assumption that it included only three arches and the assumption that it included up to eight arches.
The right panels of Figure~\ref{fig:amplitude_histograms} show the amplitudes recovered in each of these cases.
Of note is the fact that for the eight arch case all eight arches have amplitude posteriors inconsistent with amplitude of zero, which in conjunction with the injection validity tests shows that there is strong evidence that at least eight arches are present in the data, despite the fact that only six may be clearly seen in the spectrogram. 
Also of note is that for the subset of arches considered, the three arch model is in agreement with the eight arch model, such that while subtraction under the three arch assumption may be incomplete, it is not biased. 
The subtraction in the spectrogram is somewhat less rosy; while fairly effective, artifacts do remain, especially on the right edge of the arch.
This may be traced to asymmetry within the scattering mechanism, which causes a breakdown of the model's assumption and validity in this regime.
%\rhiannon{Show the plot that supports this assumption?. Investigating the inclusion of this asymmetry in the model would be a useful extension of this work.}

%\subsection{Applications to real data }

As previously mentioned, due to their width and frequency, scattered light glitches are the most common glitches to overlap true gravitational-wave signals. 
Although their morphology is distinct from morphology of the types of gravitational-wave signals that have been detected so far, meaning they are less likely to be misidentified as true gravitational waves than other signals ~\cite{Cabero:2019orq,LIGO:2021ppb,TheLIGOScientific:2017lwt,Nitz:2017lco,Cabero:2019orq,Davis:2020nyf,LIGO:2021ppb,Ashton:2021tvz}, they nonetheless may bias parameter estimation due to the inclusion of excess power in the analysis band~\cite{Macas:2022afm,Davis:2022ird}. 
Modeled subtraction of glitch power allows for the cleaning of data near or on top of the signal, without requiring assumptions about the relationship between the observed strain data and a sensor witnessing the source of the glitch, especially with regards to the amplitudes of the harmonics.
Joint inference of the glitch model and the gravitational-wave source properties, which we hope to investigate in a future work, will be required for confident subtraction of arches coincident with signals, but inference with restricted priors may still allow for glitch subtraction without affecting the underlying signal.
Particularly useful is the modelled test for the presence or absence of arches not seen in spectrograms.
This allows for checks of the effect of cleaning procedures, ensuring that there is no residual power overlapping the signal being analyzed. 

\begin{figure*}[t]
\includegraphics[width=0.75\textwidth]{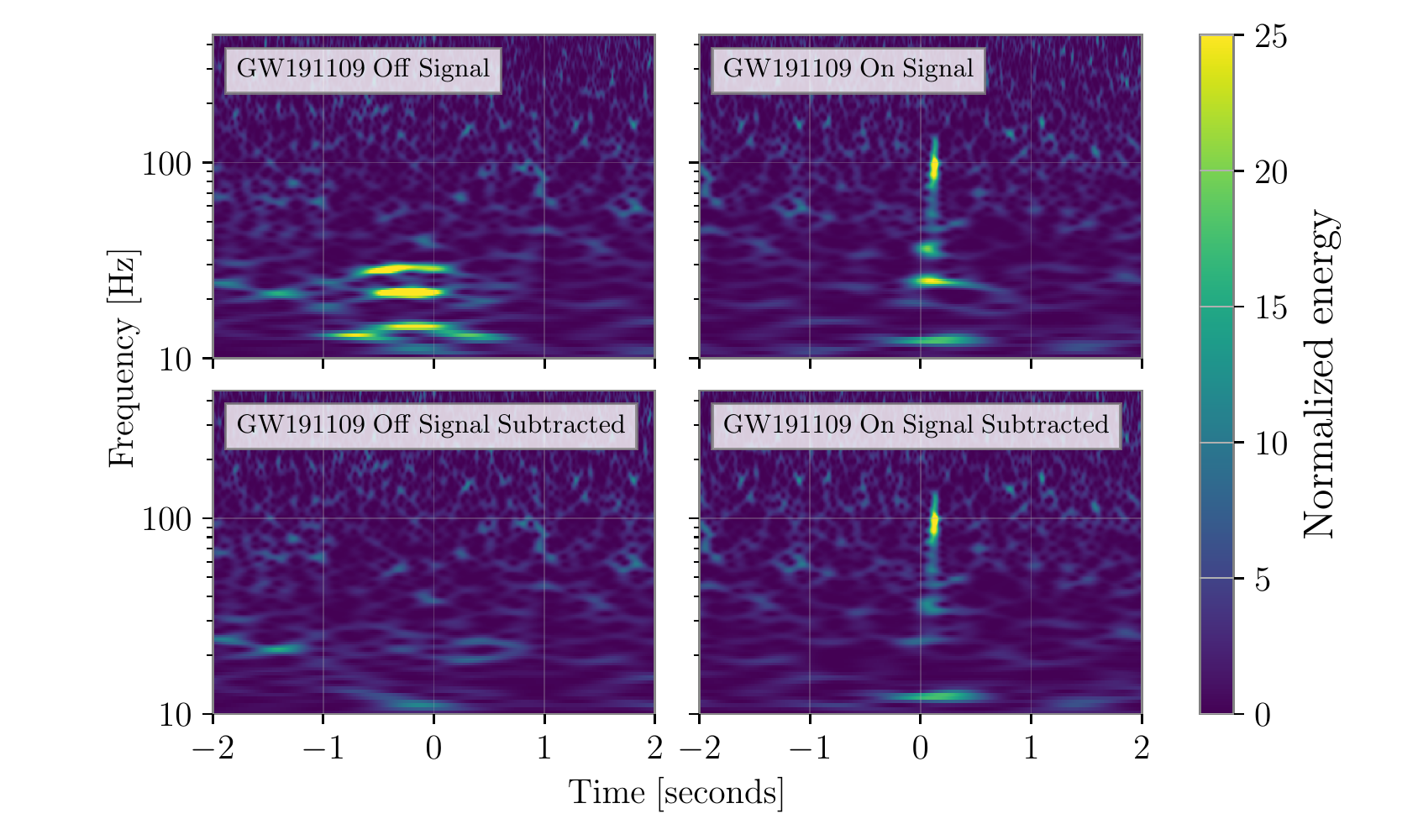}
\caption{
Spectrograms for the stretch of data around GW191109\_010717~\cite{GWTC-3}.
On the top left is data corresponding to a scattering event approximately three seconds before the gravitational wave signal. 
Inference was performed on this event, and the maximum likelihood parameters were subtracting, producing the residual plotted in the bottom left panel.
Notably, though inference was done with a model which allowed up to eight arches, the three which were visible were the only ones with amplitudes constrained away from zero.
The top right shows the stretch of data centered on the gravitational-wave signal, which overlapped with another scattering event.
For this stretch of time we perform the analysis with a three arch model, to prevent erroneous inclusion of gravitational-wave signal power in the glitch model.
Subtraction of the maximum likelihood parameters for this three arch model is shown in the bottom right panel. 
}
\label{fig:191109}
\end{figure*}

To consider this, we investigate the case of GW191109\_010717, a signal in O3~\cite{GWTC-3}. 
GW191109\_010717 was a high-mass system observed in LIGO Hanford and LIGO Livingston, which were both experiencing scattered light glitches at the time of the event. Scattered light glitches were subtracted from the data from both detectors using the Bayeswave algorithm~\cite{GWTC-3, Cornish:2020dwh}.
In LIGO Livingston the signal overlapped directly with a slow scattering arch, while in LIGO Hanford it did not, and so we focus on the case of Livingston.
GW191109\_010717 is also notable for the inference of a negative value of the effective aligned spin parameter $\chi_\text{eff}$~\cite{Ajith:2009bn,Santamaria:2010yb}, which has been tentatively connected to whether or not data that contains a scattered light glitch is used in the analysis~\cite{Davis:2022ird}. 
For the data from the LIGO Livingston detector, we apply our model in two cases: a set of arches a few seconds before the signal but not overlapping it, and the arches which directly overlap the signal.
In the off-signal case, we use an eight arch model, while for inference on the signal we restrict ourselves to 3 arch analysis.
When doing analysis of an overlapping gravitational-wave signal and scattering event, overly broad priors may allow the gravitational-wave signal power to be conflated with glitch power.
Targeted priors reduce the impact of this, since the morphology of the arches and of the gravitational-wave signal - especially the values of $f_{mod}$ which would be inferred for each - are inconsistent, but as a matter of safety we restrict ourselves to the region below $40Hz$, where the glitch has been shown to have the most effect~\cite{Davis:2022ird}.
Subtraction for each case may be seen in Figure~\ref{fig:191109}, and is visually successful.
In the off signal case, we are able to achieve successful subtraction, and notably only find support for the lower three arches which are visible.
Because the same scatterer generates the arches on-signal, this lends support to the possibility that there are only three arches in the on-signal case as well, but this cannot be confirmed for lack of joint inference.
Meanwhile, we are able to cleanly subtract the three arches we do search for in the on-signal case. 
While it is interesting to investigate how the use of this method impacts estimates of the signal's source properties, we defer this question to future studies.  

%\subsection{Conclusions}
%Using a physically motivated model for scattering glitches in gravitational-wave detectors, we perform modeled Bayesian inference of these glitches.
Modelling scattered light glitches using Bayesian inferences and a physically motivated model as presented in this letter has a number of benefits as compared to previous methods used to characterize and subtract these glitches. 
Firstly, modeled inference allows for robust subtraction of these glitches from detector data, including when coincident with true gravitational-wave signals.
Secondly, we have shown that this modeled inference provides as meaningful test of the presence or absence of higher arch harmonics which may be present in the data, including arches which are not visible within spectrograms and thus might otherwise escape detection.
Characterizing scattered light glitches in this way is independent of whether a source of the scattered light has been identified. 
Together, these two features have the potential to make this an important tool in mitigation of scattered light glitches in the future. 
Because our method uses the standard inference tools of gravitational wave parameter estimation, there are good prospects for future development of joint inference methods, allowing for disentanglement of scattered light glitches from true signals, and reducing the chances of misidentifying scattered light with features in these signals.

%\end{document} % uncomment this to get a sense of the word count without the acknowledgements

\begin{acknowledgments}
The authors thank the LIGO-Virgo-KAGRA Detector Characterization and Parameter Estimation groups for their input and suggestions during the development of this wor.
The authors thank Arthur Tolley, Ian Harry, Andrew Lundgren, Gareth Cabourn Davies, Colm Talbot, Sophie Hourihane, and Robert Schofield for productive discussions, and also thank Colm Talbot for access to the extended Dynesty sampling methods used in this work.
We also thank Andrew Lundgren for his comments during internal review of this paper.
DD and RPU are supported by the NSF as a part of the LIGO Laboratory.

This material is based upon work supported by NSF’s LIGO Laboratory 
which is a major facility fully funded by the 
National Science Foundation.
LIGO was constructed by the California Institute of Technology 
and Massachusetts Institute of Technology with funding from 
the National Science Foundation, 
and operates under cooperative agreement PHY-1764464. 
Advanced LIGO was built under award PHY-0823459.
The authors are grateful for computational resources provided by the 
LIGO Laboratory and supported by 
National Science Foundation Grants PHY-0757058 and PHY-0823459.
This work carries LIGO document number P2200350.
\end{acknowledgments}

\section*{Data Availability Statement}

The data that support the findings of this study are available from the corresponding author upon reasonable request.

%\nocite{*}
\bibliography{main}% Produces the bibliography via BibTeX.

\clearpage
\end{document}